\documentclass[aps,twocolumn,showpacs,preprintnumbers,amsmath,amssymb,showkeys,floatfix,superscriptaddress]{revtex4}

\usepackage{graphicx,epsf,epsfig}
\usepackage{dcolumn}
\usepackage{bm}
\usepackage{color}
\usepackage{ulem}

\begin{document}


\title{Raman study of Fano interference in \textit{p}-type doped silicon}
\author{Brian G. Burke}
\author{Jack Chan}
\author{Keith A. Williams}
\email[Corresponding Author: ]{kwilliams@virginia.edu}
\affiliation{Department of Physics, University of Virginia, Charlottesville, Virginia 22904}
\author{Zili Wu}
\affiliation{Chemical Sciences Division, Center for Nanophase Materials Sciences, Oak Ridge National Laboratory, Oak Ridge, Tennessee, 37831}
\author{Alexander A. Puretzky}
\author{David B. Geohegan}
\affiliation{Materials Science and Technology Division, Center for Nanophase Materials Sciences, Oak Ridge National Laboratory, Oak Ridge, Tennessee 37831}

\begin{abstract}
As the silicon industry continues to push the limits of device dimensions, tools such as Raman spectroscopy are ideal to analyze and characterize the doped silicon channels. The effect of inter-valence band transitions on the zone center optical phonon in heavily \textit{p}-type doped silicon is studied by Raman spectroscopy for a wide range of excitation wavelengths extending from the red (632.8 nm) into the ultra-violet (325 nm). The asymmetry in the one-phonon Raman lineshape is attributed to a Fano interference involving the overlap of a continuum of electronic excitations with a discrete phonon state. We identify a transition above and below the one-dimensional critical point (E$_{\Gamma_{1}}$ = 3.4 eV) in the electronic excitation spectrum of silicon. The relationship between the anisotropic silicon band structure and the penetration depth is discussed in the context of possible device applications.
\end{abstract}

\pacs{Valid PACS appear here}
\maketitle

\section{Introduction}
As the silicon industry continues to push the limits of device dimensions, tools such as Raman spectroscopy are ideal to analyze and characterize the doped silicon channels being developed. Due to the small penetration depth ($\sim$ 8 nm) of 325 nm light in silicon and motivated by recent doped silicon-on-insulator (SOI) devices, confocal micro-Raman spectroscopy offers a unique method to probe both bulk and surface scattering. Raman scattering in solids can arise through a number of different processes including scattering by phonons, plasmons, intraband, and interband electronic transitions. Intraband electronic Raman scattering, where the initial and final states are in the same band and the intermediate state is in a different band, has been thoroughly investigated both experimentally and theoretically \cite{mills}. Only a few experiments have been done on interband electronic Raman scattering and research has focused on inter-valence band scattering in \textit{p}-type doped silicon \cite{klein,pinczuk,cerdeira,cerdeira2}.

Experiments were conducted to investigate the effect of free holes on the optical phonons of heavily \textit{p}-type doped silicon as well as an asymmetry in the Raman lineshape of the zone center optical phonon (521 cm$^{-1}$) \cite{cerdeira,cerdeira2}. When exposed to incoming light, the doped silicon produces a continuum of inter-valence band electronic excitations, which overlaps the Raman-active one-phonon energy. As a result, the overlap between the electronic continuum and discrete state causes interference effects and a phenomenon termed Fano interference arises. Fano interference was first described by U. Fano after observing an asymmetry in the electron energy loss spectrum of helium \cite{fano}. The asymmetry was attributed to a quantum interference involving the overlap of the wavefunctions of discrete states and a continuum. The helium lineshape was fitted with a Fano function, which depends on two parameters. In this paper, we present a detailed experimental and theoretical study of the Raman spectrum of \textit{p}-type doped silicon for a range of carrier concentrations and a wider array of excitation wavelengths, extending into the ultra-violet (UV), to probe the surface and bulk silicon states.

\section{Theory}
The Fano effect in \textit{p}-type silicon is associated with the resonant interaction of a discrete state with a continuum (Fig. 1). The first state, being a discrete state, represents a hole in its acceptor ground state and one locally excited optical phonon. The second state, being a continuum, represents no phonons being created and a hole excited into the p$_{3/2}$ valence band \cite{gajic}. The Fano effect in \textit{p}-type silicon has been observed in photoconductivity and Raman spectroscopy experiments \cite{cerdeira2,gajic}. The asymmetry due to the zone center ($\vec{k}$ = 0) optical phonon changes as a function of temperature, excitation wavelength, and doping concentration \cite{cerdeira2}. Interaction solely with the hole continuum is due to the fact that shallow acceptor levels are separated by energies comparable to phonons.

\begin{figure}[h!]
\hskip 0.2cm\centerline{\epsfxsize=3.5in\epsfbox{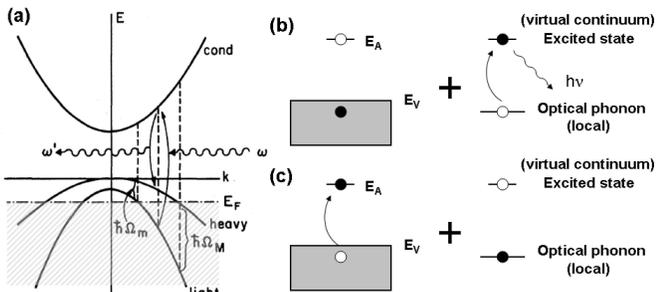}}
\vskip -0.4cm
\caption{(a) Isotropic parabolic energy bands with the Fermi energy E$_{F}$ in the valence bands \cite{kanehisa}. Diagram of Fano interference: (b) Discrete state; represents a hole in its acceptor ground state and one locally excited optical phonon (c) Continuum; represents no phonons being created and a hole excited into the p$_{3/2}$ valence band.}
\label{fig1}
\end{figure}

The changes in the Raman lineshape are due to the existence of a continuum of electronic excitations produced by inter-valence band transitions whose energy overlaps that of the zone center optical phonon. As the boron doping concentration is increased in silicon, the light and heavy hole bands are split further from one another. One-electron excitations, which are infrared forbidden but Raman allowed at $\vec{k}$ = 0, scatter electromagnetic radiation which is observed as Raman scattering \cite{cerdeira2}. Using a modified Fano function, 
\begin{equation}
F(q,\Gamma) = \frac{(q + \epsilon)^{2}}{1 + \epsilon^{2}}	
\label{chp3eq1}
\end{equation}
where $\epsilon = \frac{(\omega - \Omega)}{\Gamma}$, Cardona \textit{et al}. were able to fit the experimental Raman data of \textit{p}-type doped silicon and extract the two parameters $q$ and $\Gamma$ \cite{cerdeira2}. $\Gamma$ is the squared matrix element of the coupling between the continuum and the discrete phonon state, which depends only on the carrier concentration, while $q$ squared represents the ratio of the scattering probability of the discrete state to that of the continuum, and depends on both the carrier concentration and excitation wavelength \cite{cerdeira}.

The peak position of the intrinsic one-phonon Raman peak is shifted as the doping concentration is increased. The doped silicon critical frequency ($\Omega$) is related to the phonon frequency by \cite{cerdeira2}
\begin{equation}
\Omega = \Omega_{int}	+ \delta\Omega
\label{chp3eq2}
\end{equation}
where $\Omega_{int}$ represents the intrinsic silicon peaks and $\delta\Omega$ represents the shift of the critical frequency from the intrinsic frequency. The shift is related to the parameters $q$ and $\Gamma$ by \cite{cerdeira2}
\begin{equation}
\delta\Omega = \delta\Omega_{max}	- \frac{\Gamma}{q}
\label{chp3eq3}
\end{equation}
where $\delta\Omega_{max}$ represents the shift in the position of the maximum peak of the doped silicon peak from the intrinsic silicon peak.

Crystalline semiconductors exhibit first-order Raman scattering only by phonons with $\vec{k}$ $\cong$ 0 \cite{cardona}. Since the photon wave vector ($k_{photon}$ = 2$\pi$/$\lambda$) is much less than the Brillouin-zone size (2$\pi$/$a_{0}$), one can assume vertical transitions and set q = 0 and satisfy conservation of crystal momentum ($k_{s}$ = $k_{l}$ $\pm$ q) \cite{ashcroft}. Incoming light of a particular wavelength ($\omega$) excites electrons in the light hole band to the conduction band, which then interact with a phonon, emit a scattered photon ($\omega$') and de-excite into the heavy hole band. Assuming isotropic parabolic energy bands, these inter-valence band transitions give a Raman spectrum with cutoffs at $\Omega_{m}$ and $\Omega_{M}$, representing the minimum and maximum energies defined by the valence band curvature and the Fermi level for a given carrier concentration (Fig. 1) \cite{kanehisa}. However, silicon has highly anisotropic valence bands and in order to obtain quantitative results it is necessary to have knowledge of the detailed band structure over the whole Brillouin zone.

The Fermi level is not dependent on the excitation wavelength. As the excitation energy is varied, corresponding to different energy gaps for the inter-valence band transitions, different orientations in \textit{k}-space will have higher Raman efficiencies. Doping may also shift the valence and conduction bands vertically as well as the Fermi level. For each excitation wavelength, the 3D $E-k$ plot for silicon should be analyzed to determine which crystal orientation will dominate and which light-heavy valence band transitions will contribute to the Raman spectrum. For certain crystal orientations, such as [110], the minimum energy $\Omega_{m}$ is no longer less than the zone center optical phonon ($\Omega_{m}$ $>$ $\Omega$ = 65 meV), which disallows the overlap of the continuum with the discrete phonon state \cite{cerdeira2}.

\section{Experiment and Results}
Raman data was collected from a triple-axis spectrometer (Jobin Yvon Horiba, T64000) using a liquid nitrogen cooled CCD detector in the back-scattering configuration. Measurements were performed with several cw lines of a HeNe (632.8 nm), diode (532 nm), and HeCd (442 nm and 325 nm) laser and the second harmonic of a pulsed Ti:sapphire laser (Mira 900, Coherent, 5 ps pulses, 76 MHz repetition rate, 382.6 nm). The excitation light was focused onto a sample using a long distance objective (Nikon, 50x/0.45) to a spot size of $\sim$ 1 $\mu$m. The carrier concentrations for the silicon samples were determined by sheet resistance and Hall Probe measurements.

\subsection{Si(100) Samples}
Measurements were performed at room temperature on thermally diffused boron doped Si(100) wafers, which were cleaned with a buffered oxide etch (BOE) and dried with N$_{2}$. The Raman data obtained for the three thermally diffused boron doped samples are shown in Figures 2 and 3. The experimental data (discrete points) were then fitted with the modified Fano function (solid lines) with a least-squares computer fit which used $q$ and $\Gamma$ as adjustable parameters.

\begin{figure}[h!]
\hskip 0.2cm\centerline{\epsfxsize=3.2in\epsfbox{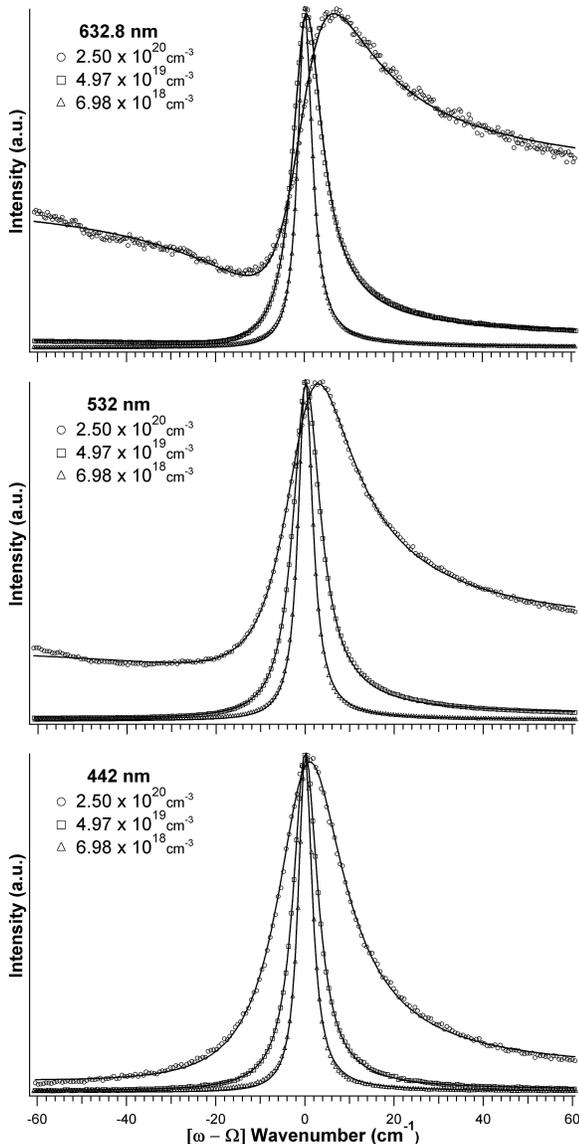}}
\vskip -0.4cm
\caption{Raman lineshapes vs. scattering wavelength for three thermally diffused boron doped Si(100) samples at excitation wavelengths 632.8 nm, 532 nm, and 442 nm (1800 gr/mm). The experimental data (discrete points) were then fitted with the modified Fano function (solid lines) with a least-squares computer fit which used $q$ and $\Gamma$ as adjustable parameters. The spectra were normalized by the maximum peak value of each concentration.}
\label{fig2}
\end{figure}

\begin{figure}[h!]
\hskip 0.2cm\centerline{\epsfxsize=3.2in\epsfbox{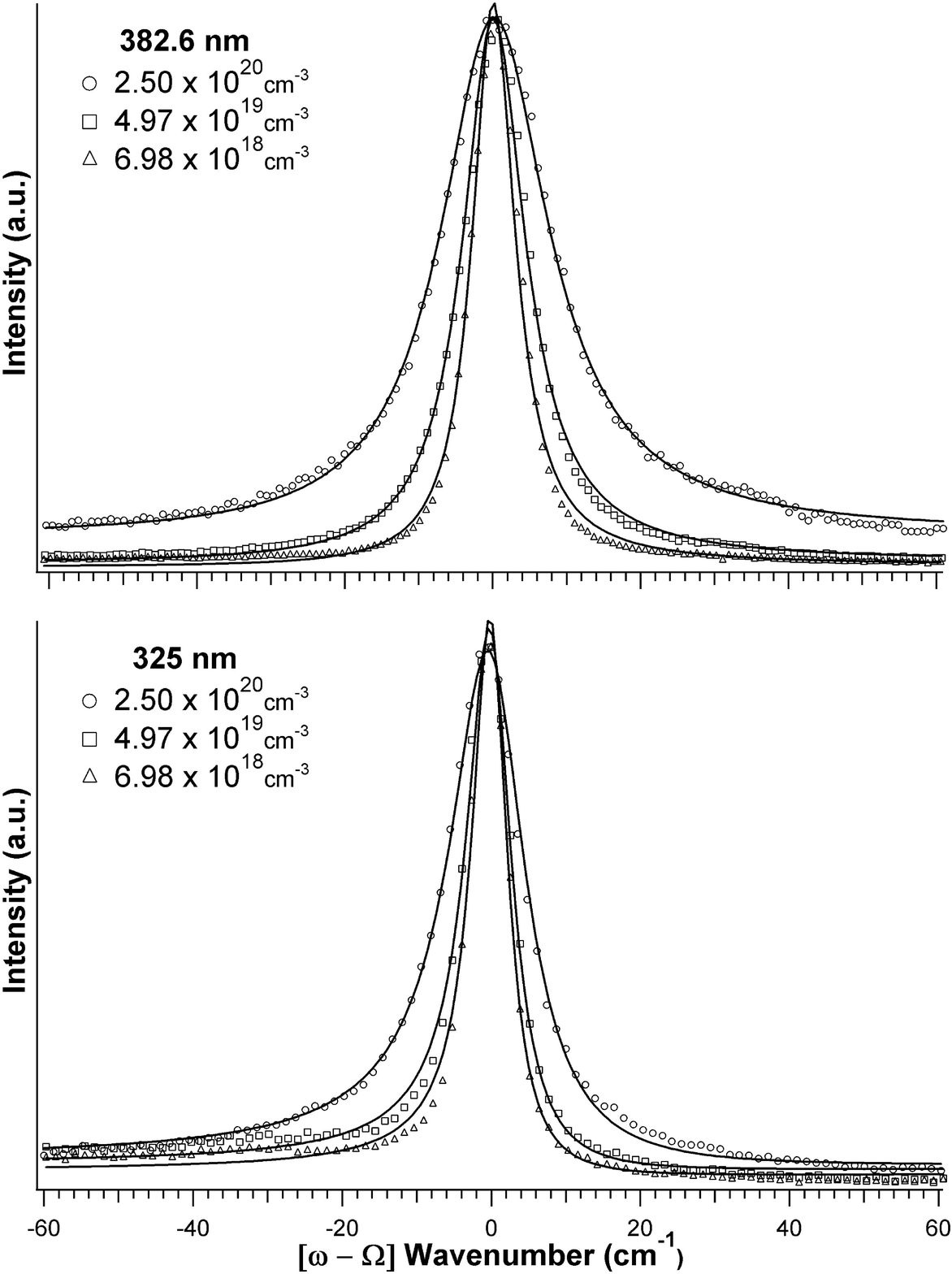}}
\vskip -0.4cm
\caption{Raman lineshapes vs. scattering wavelength for three thermally diffused boron doped Si(100) samples at excitation wavelengths 382.6 nm and 325 nm (2400 gr/mm). The experimental data (discrete points) were then fitted with the modified Fano function (solid lines) with a least-squares computer fit which used $q$ and $\Gamma$ as adjustable parameters. The spectra were normalized by the maximum peak value of each concentration.}
\label{fig3}
\end{figure}

The values extracted from the fit for $\Gamma$, $q$, $\delta\Omega$, and $\delta\Omega_{max}$ are listed in Tables I and II. The fit obtained for Fig. 2 has excellent agreement ($<$ 2$\%$ error), whereas the fitting function begins to deviate from the experimental values as the excitation wavelength moves deeper into the UV.

\begin{table}[ht]
\caption{Parameters obtained by fitting experimental first-order Raman spectra with the modified Fano function (\ref{chp3eq1}). Spectra were obtained for three different excitation wavelengths (632.8 nm, 532 nm, and 442 nm).}
\centering
\begin{tabular}{c c c c c c c}
\hline
Doped Si (cm$^{-3}$) & \multicolumn{3}{c}{$\Gamma$ (cm$^{-1}$)} & \multicolumn{3}{c}{$q$} \\
Concentration & 632.8 & 532 & 442 & 632.8 & 532 & 442 \\
\hline\hline
2.50 $\times$ 10$^{20}$ & 9.22 & 9.39 & 9.12 & 1.39 & 3.13 & 8.01 \\
4.97 $\times$ 10$^{19}$ & 4.05 & 3.79 & 3.33 & 7.71 & 14.6 & 35.4 \\
6.98 $\times$ 10$^{18}$ & 1.93 & 1.98 & 1.87 & 23.6 & 37.3 & 32.2 \\
\hline
\end{tabular}
\\
\vspace{0.25cm}
\begin{tabular}{c c c c c c c}
\hline
Doped Si (cm$^{-3}$) & \multicolumn{3}{c}{$\delta\Omega$ (cm$^{-1}$)} & \multicolumn{3}{c}{$\delta\Omega_{max}$ (cm$^{-1}$)} \\
Concentration & 632.8 & 532 & 442 & 632.8 & 532 & 442 \\
\hline\hline
2.50 $\times$ 10$^{20}$ & -7.8 & -7.8 & -5.8 & -1.17 & -4.8 & -4.66 \\
4.97 $\times$ 10$^{19}$ & -0.8 & -2.0 & -0.3 & -0.27 & -1.74 & -0.21 \\
6.98 $\times$ 10$^{18}$ & -0.3 & -1.6 & -0.6 & -0.22 & -1.55 & -0.54 \\
\hline
\end{tabular}
\label{chp3tab1}
\end{table}

\begin{table}[ht]
\caption{Parameters obtained by fitting experimental first-order Raman spectra with the modified Fano function (\ref{chp3eq1}). Spectra were obtained for two different excitation wavelengths (382.6 nm and 325 nm).}
\centering
\begin{tabular}{c c c c c}
\hline
Doped Si (cm$^{-3}$) & \multicolumn{2}{c}{$\Gamma$ (cm$^{-1}$)} & \multicolumn{2}{c}{$q$} \\
Concentration & 382.6 & 325 & 382.6 & 325 \\
\hline\hline
2.50 $\times$ 10$^{20}$ & 8.86 & 6.19 & 45.4 & -13.2 \\
4.97 $\times$ 10$^{19}$ & 5.15 & 3.66 & 46.4 & -13.2 \\
6.98 $\times$ 10$^{18}$ & 3.50 & 2.89 & 39.4 & -15.0 \\
\hline
\end{tabular}
\\
\vspace{0.25cm}
\begin{tabular}{c c c c c}
\hline
Doped Si (cm$^{-3}$) & \multicolumn{2}{c}{$\delta\Omega$ (cm$^{-1}$)} & \multicolumn{2}{c}{$\delta\Omega_{max}$ (cm$^{-1}$)} \\
Concentration & 382.6 & 325 & 382.6 & 325 \\
\hline\hline
2.50 $\times$ 10$^{20}$ & -3.6 & -1.8 & -3.4 & -2.27 \\
4.97 $\times$ 10$^{19}$ & -0.3 & -0.8 & -0.19 & -1.08 \\
6.98 $\times$ 10$^{18}$ & 0.2 & -0.8 & 0.29 & -0.99 \\
\hline
\end{tabular}
\label{chp3tab2}
\end{table}

As the excitation wavelength decreases towards the UV, the $\Gamma$ values are no longer comparable and the $q$ values become increasingly higher and then negative. If $\Gamma$ is held constant, which according to Cardona \textit{et al}. should be independent of the excitation wavelength, for a doped silicon sample with carrier concentration 2.50 $\times$ 10$^{20}$ cm$^{-3}$, the experimental data can be fitted as a function of $q$ only and compared to the different excitation wavelengths (Fig. 4).

\begin{figure}[h!]
\hskip 0.2cm\centerline{\epsfxsize=3.4in\epsfbox{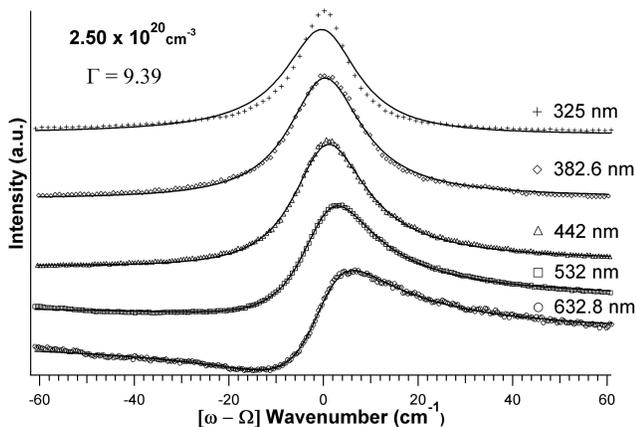}}
\vskip -0.4cm
\caption{Raman lineshapes vs. scattering wavelength for one thermally diffused boron doped Si(100) sample with carrier concentration 2.50 $\times$ 10$^{20}$ cm$^{-3}$. The experimental data (discrete points) were then fitted with the modified Fano function (solid lines) with a least-squares computer fit which used $\Gamma$ = 9.39 cm$^{-1}$ and $q$ as an adjustable parameter. The spectra have been shifted vertically with respect to one another.}
\label{fig4}
\end{figure}

\begin{table}[ht]
\caption{Parameters obtained by fitting experimental first-order Raman spectra with the modified Fano function (\ref{chp3eq1}) and holding $\Gamma$ = 9.39 cm$^{-1}$. Spectra were obtained for a single carrier concentration of 2.50 $\times$ 10$^{20}$ cm$^{-3}$.}
\centering
\begin{tabular}{c c c c}
\hline
Excitation Wavelength (nm) & $q$ & $\delta\Omega$ (cm$^{-1}$) & $\delta\Omega_{max}$ (cm$^{-1}$) \\
\hline\hline
632.8 & 1.47 & -7.8 & -1.41 \\
532 & 3.13 & -7.8 & -4.80 \\
442 & 8.05 & -5.8 & -4.63 \\
382.6 & 27.1 & -3.8 & -3.45 \\
325 & -23.9 & -1.8 & -2.19 \\
\hline
\end{tabular}
\label{chp3tab3}
\end{table}

The values extracted from the fit for $q$, $\delta\Omega$, and $\delta\Omega_{max}$ are listed in Table III. The theoretical fit has excellent agreement for all values down to the 382.6 nm excitation, but does not converge for the 325 nm excitation. Also, the $q$ values are negative for the 325 nm excitation. The interpretation of these results will be addressed below.

\subsection{SOI Samples}
Surface molecular doping of intrinsic silicon-on-insulator (SOI) Si(100) samples has been investigated and molecular monolayers on hydrogen terminated silicon for various thicknesses (20 $-$ 200 nm) have been formed. Two molecular dopant precursors have been utilized; allyboronic acid pinacol ester (boronic acid), which forms a covalent Si-C bond, and decaborane (B$_{10}$H$_{14}$), which forms a hydrogen Si-H bond \cite{ho,chen}. The samples underwent rapid thermal annealing (RTA) for various temperatures and time scales and the carrier concentrations were then calculated by sheet resistance measurements. RTA is a manufacturing process which heats silicon samples to high temperatures ($\sim$ 1100$^{\circ}$C) on a timescale of several seconds. During RTA, the Si-C or Si-H bonds are broken and the remaining boron atoms become active and diffuse into the surface of the silicon.

Two SOI samples (75 nm and 25 nm thickness) which were doped with the boronic acid precursor and underwent RTA (1050$^{\circ}$C, 500 s) produced the highest carrier concentrations. The calculated carrier concentrations were 1.54 $\times$ 10$^{21}$ cm$^{-3}$ and 5.20 $\times$ 10$^{21}$ cm$^{-3}$ respectively. The measured intrinsic SOI (75 nm thickness) carrier concentration was 8.63 $\times$ 10$^{17}$ cm$^{-3}$.

\begin{figure}[h!]
\hskip 0.2cm\centerline{\epsfxsize=3.4in\epsfbox{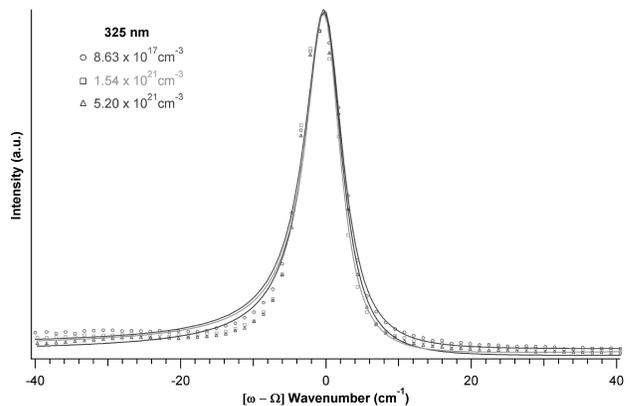}}
\vskip -0.4cm
\caption{Raman lineshapes vs. scattering wavelength for three molecularly doped SOI Si(100) samples with different carrier concentrations. The experimental data (discrete points) were then fitted with the modified Fano function (solid lines) with a least-squares computer fit which used $\Gamma$ and $q$ as adjustable parameters. There is no significant change, regardless of the carrier concentration.}
\label{fig5}
\end{figure}

The Raman data obtained for the three molecularly doped samples are shown in Fig. 5. The experimental data (discrete points) were then fitted with the modified Fano function (solid lines) with a least-squares computer fit which used $q$ and $\Gamma$ as adjustable parameters. No significant changes in the Raman lineshape were observed, despite the large difference in measured carrier concentration.

\section{Discussion}
The parameter $q$ has a strong dependence on the excitation wavelength as can be seen in Fig. 4. However, a transition can be observed in the Raman lineshape between the 382.6 nm and 325 nm excitation wavelengths. The Raman lineshape for the 325 nm excitation, with $\Gamma$ held constant, deviates from the theory proposed by Cardona \textit{et al}. \cite{cerdeira2}.

One explanation for the divergence is that the carrier concentration is inhomogeneous. The inhomogeneity is not uncommon at such high carrier concentrations, which are close to the solubility limit, and would cause light with different penetration depths to probe regions of different Fermi energies \cite{cerdeira2}. However, this explanation seems unlikely due to the fact that the other two thermally diffused boron silicon samples, which are at lower carrier concentrations, also show this deviation from theory and a transition between the 382.6 nm and 325 nm excitation wavelengths. Additionally, all molecular doped SOI samples showed no change in the Raman lineshape for the 325 nm excitation, despite having the highest carrier concentrations. These results may indicate a transition in the silicon band structure from bulk to surface bands.

A second possible explanation for the divergence is to consider the penetration depth and electronic excitation spectrum for silicon (Fig. 6) \cite{hull,globus,aspens,herzinger,loewenstein,hecht}. The previous work done by Cardona \textit{et al}. can be considered an investigation of the micron region (450 nm $-$ 650 nm), whereas the present work considers the submicron region (325 nm $-$ 632.8 nm). The approximation made by Cardona \textit{et al}. of the combined density of states for the continuum of inter-valence band transitions as one-dimensional critical points at E$_{\Gamma_{1}}$ = 3.4 eV and E$_{\Gamma_{2}}$ = 4.2 eV leads to a semi-quantitative comparison, since those expressions are only valid in the immediate neighborhood of the resonant gap \cite{cerdeira,cerdeira2}.

\begin{figure}[h!]
\hskip 0.2cm\centerline{\epsfxsize=3.4in\epsfbox{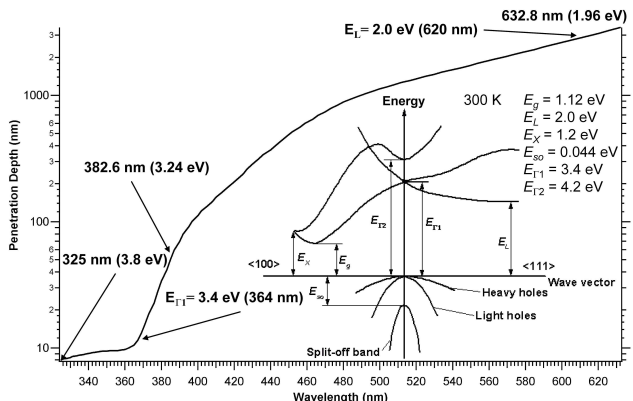}}
\vskip -0.4cm
\caption{Penetration depth vs. wavelength and electronic excitation spectrum of silicon (inset adapted from \cite{ioffe}). An inflection point in the logarithmic plot of the penetration depth can be seen at 364 nm. This point is equal to 3.4 eV in energy, which corresponds to the first critical $\Gamma$ point (E$_{\Gamma_{1}}$) at $\vec{k}$ = 0 in the band structure of silicon. The 382.6 nm (3.24 eV) excitation is below the critical point and follows Cardona's theory, while the 325 nm (3.8 eV) excitation is above the critical point and enters a new regime of Raman scattering.}
\label{fig6}
\end{figure}

At higher energies, greater than the first critical point (E$_{\Gamma_{1}}$ = 3.4 eV), one must consider the anisotropic valence bands and calculate the detailed band structure over the whole Brillouin zone via a 3D $E-k$ plot to account for the inter-valence band transitions that will play a role in the Raman scattering.

By converting the energies of the critical points to excitation wavelengths, an inflection point in the logarithmic plot of the penetration depth in silicon can be seen at 364 nm. This point is equal to 3.4 eV in energy, which corresponds to the first critical $\Gamma$ point (E$_{\Gamma_{1}}$) at $\vec{k}$ = 0 in the band structure of silicon. The 382.6 nm (3.24 eV) excitation is below the critical point and follows the proposed theory by Cardona \textit{et al}., while the 325 nm (3.8 eV) excitation is above the critical point and enters a \textit{new regime} of Raman scattering.

\begin{figure}[h!]
\hskip 0.2cm\centerline{\epsfxsize=3.2in\epsfbox{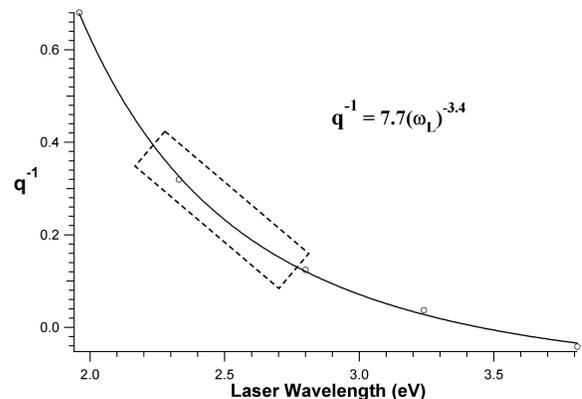}}
\vskip -0.4cm
\caption{Plot of q$^{-1}$ vs. excitation wavelength for thermally diffused boron silicon sample with carrier concentration 2.50 $\times$ 10$^{20}$ cm$^{-3}$. A power law dependence is observed, which deviates from the linear dependence prediction of Cardona \textit{et al}. (noted in outlined box).}
\label{fig7}
\end{figure}

Additionally, Cardona \textit{et al}. found a linear dependence in the plot of q$^{-1}$ vs. excitation wavelength for a heavily \textit{p}-type doped silicon sample (1.6 $\times$ 10$^{20}$ cm$^{-3}$), whereas a power law dependence is found for a thermally diffused boron silicon sample with carrier concentration 2.50 $\times$ 10$^{20}$ cm$^{-3}$ (Fig. 7) \cite{cerdeira}. Cardona \textit{et al}. probed a smaller range of experimental data points, which would appear to be linear. However, when analyzing the full spectrum, the dependence is clearly non-linear.

\section{Conclusions}
Based on the experimental and theoretical analysis by Raman spectroscopy, for a wide range of excitation wavelengths (325 $-$ 632.8 nm), we have studied the effect of inter-valence band transitions on the zone center optical phonon and observed Fano interference in heavily p-type doped silicon. We were able to fit the asymmetry in the one-phonon Raman lineshape by utilizing a modified Fano function and have identified a transition above and below the one-dimensional critical point (E$_{\Gamma_{1}}$ = 3.4 eV) in the electronic excitation spectrum of silicon. By analyzing the logarithmic plot of the penetration depth in silicon vs. excitation wavelength, we have identified an inflection point which corresponds to the first critical point E$_{\Gamma_{1}}$ and a change in the inter-valence band transitions. Energies higher than the first critical point cause the system to enter into a new regime of Raman scattering. As doped silicon devices are scaled down in size, Raman spectroscopy is an important tool to study both bulk and surface scattering.

\section{Acknowledgements}
The authors would like to thank Lloyd Harriott, Smitha Vasudevan and Avik Ghosh for helpful discussions and Wenjing Yin for her assistance with Hall Probe measurements. A portion of this research at Oak Ridge National Laboratory's Center for Nanophase Materials Science was sponsored by the Scientific User Facilities Division, Office of Basic Energy Sciences, U.S. Department of Energy.

\end{document}